%% file: paper.tex
\input{mak.tex}
\documentclass[12pt,epsfig]{iopart}
\usepackage{epsfig}
\begin{document}

\title{Block recursion and Green matrices}

\author{Kamal Krishna Saha\footnote{Email : kamal@bose.res.in} and Abhijit Mookerjee}
\address{ S. N. Bose National Centre for Basic Sciences. Block-JD, Sector-III, \\ 
Salt Lake City, Kolkata-700098, India.}

\begin{abstract}
{We present here generalization of the recursion method of Haydock \etal \cite{hhk} for the
calculation of Green matrices (in angular momentum space). Earlier approaches
concentrated on the diagonal elements, since the focus was on spectral densities. However,
calculations of configuration averaged response functions or neutron scattering
cross-sections require the entire Green matrices and self-energy matrices obtained from it.
This necessitated the generalization of the recursion method presented here with examples.} 
\end{abstract}

\date{\today}
\parindent 0pt

\section{Introduction}

The augmented space recursion  carried out in a minimal basis set representation 
of the tight-binding linear muffin-tin orbitals method (TB-LMTO-ASR) has been proposed earlier
by us \cite{asr,asr2} as  technique for the incorporation of the effects of
configuration fluctuations 
 for random substitutionally disordered alloys.
 This can be achieved without the usual problems of violation
of the Herglotz analytic properties \cite{her} of the approximated configuration averaged Green functions
for the Schr\"odinger equation for these random alloys. Although our initial focus was
with configuration averages of density of states and spectral functions, recently we have
proposed using the TB-LMTO-ASR for the study of configuration averaged optical conductivities \cite{opt} or coherent and incoherent neutron scattering cross-sections \cite{alam}. These
calculations require the full Green matrices in angular momentum space and not only their
diagonal elements. We propose here the use of a generalization of the recursion method of Haydock \etal \cite{hhk}. The block recursion technique had been introduced earlier by Godin
and Haydock \cite{gh1,gh2} in the very different context for obtaining the scattering S-matrix for finite scatterers attached to perfect leads. We shall borrow their ideas and set up a
block recursion in angular momentum space (rather than the lead space, as in Godin and Haydock's work) in order to obtain the Green matrices (in angular momentum space) directly.

\section{Methodology}

\subsection{The TB-LMTO Hamiltonian in augmented space and the recursion method}

The augmented space recursion based on the tight-binding linear muffin-tin
orbitals method (TB-LMTO-ASR) has been described thoroughly in a series of articles \cite{Am}-\cite{Am8}.
We shall introduce the salient features of the ASR  which will be required by us in our
subsequent discussions.

 We shall start from a first principle tight-binding linear muffin-tin 
orbitals (TB-LMTO) \cite{Ander1,Ander2} in the most-localized
representation ($\alpha$ representation). This is necessary, because the subsequent
recursion requires a sparse representation of the Hamiltonian. In this 
representation, the second order alloy Hamiltonian is given by,

\[{\mathbf H}^{(2)} = {\mathbf E}_\nu + {\mathbf h} - \mathbf{hoh}  \]
where
\begin{eqnarray}
{\mathbf h}  =  \sum_{R} \left({\bf C}_{R} - {\mathbf E}_{\nu R}\right) \enskip {\cal P}_{R} + \sum_{R}\sum_{R'}{\mathbf \Delta}_{R}^{1/2} 
\ {\mathbf S}_{RR'} \ {\mathbf \Delta}_{R'}^{1/2} \enskip {\cal T}_{RR'}\nonumber\\ 
\phantom{x}\nonumber\\
{\mathbf o} = \sum_R \mathbf{o}_R \ {\cal P}_R
\label{eqa} 
\end{eqnarray}

{\bf C}$_R$, {\bf E}$_{\nu R}$ , ${\mathbf\Delta}_R$  and ${\mathbf o}_R$ are diagonal matrices in angular momentum space :
\[ {\mathbf C}_R = C_{_{RL}}\ \delta_{_{LL'}} \quad {\mathbf E}_{\nu R}= E_{_{\nu RL}}\ \delta_{_{LL'}}
\quad \mathbf{\Delta}_R = \Delta_{_{RL}}\ \delta_{_{LL'}}\ \mbox{ and } {\mathbf o}_R=o_{_{RL}}\ \delta_{_{LL'}}\]
\noindent and {\bf S}$_{RR'}$ = $S_{RL,R'L'}$ is a matrix of rank $L_{\mathrm{max}}$.
 ${\cal P}_{R} = \vert R\rangle\langle R\vert$ and ${\cal T}_{RR'} = \vert R\rangle\langle R'\vert$ are projection and transfer operators 
in the Hilbert space ${\cal H}$ spanned by the tight-binding basis $\{|R\rangle\}$. Here, $R$ refers to the position of atoms
in the solid  and $L$ is a composite label $\{\ell,m,m_s\}$ for the angular momentum quantum numbers. ${\bf C}$, ${\mathbf \Delta}$ 
and ${\mathbf o}$'s are potential parameters of the TB-LMTO method; these are diagonal matrices in the angular momentum indices, 
${\bf o}^{-1}$ has dimension of energy and ${\bf E}_\nu$'s are the energy windows about which the muffin-tin orbitals are linearized.

For a disordered binary alloy we may write : 

\begin{eqnarray}
{ C}_{RL} & = &  C_L^A\  n_R + C_L^B\ \left( 1-n_R \right) \nonumber \\
\Delta_{RL}^{1/2} & = & \left( \Delta_L^A \right)^{1/2} n_R 
+ \left(\Delta_L^B \right)^{1/2}\left( 1-n_R \right)  \nonumber \\
{o}_{RL} & = &  o_L^A\  n_R + o_L^B\ \left( 1-n_R \right)
\label{eqb} 
\end{eqnarray}

Here $\{n_R\}$ are the random site-occupation variables which take values 1 and 0 
 depending upon whether the muffin-tin labeled by $R$ is occupied by $A$ or
$B$-type of atom. The atom sitting at $\{R\}$ can either be of type $A \ (n_R=1)$ with probability $x$ or 
$B \ (n_R=0)$ with probability $y$. The augmented space formalism (ASF) now introduces the space of configurations 
of the set of binary random variables  $\{n_R\}$ : $\Phi$. 

In the absence of short-ranged order, each random variable  $n_R$ has associated with it an operator  ${\bf M}_R$
whose spectral density is its probability density :
\begin{eqnarray}
 p(n_R) &=& x \delta(n_R-1) + y\delta(n_R) \nonumber \\ 
&=& -\frac{1}{\pi} \lim_{\delta\rightarrow 0} \ \mbox{Im}  
\langle \uparrow_R|\left[(n_R+i\delta) {\mathbf I}-{\mathbf M}_R\right]^{-1}|\uparrow_R\rangle 
\label{eqc}
\end{eqnarray}

where ${\mathbf M}_R$ is an operator whose eigenvalues 1, 0 correspond to the observed values of $n_R$ and
whose corresponding eigenvectors $\{|1_R\rangle,|0_R\rangle\}$ span a configuration space
$\phi_R$ of rank 2. We may change the basis to  $\{|\uparrow_R\rangle,|\downarrow_R\rangle\}$ :
\begin{eqnarray*}
\vert\uparrow_{R}\rangle &\eq& \sqrt{x}\ \vert {0_{R}} \rangle + \sqrt{y}\ \vert {1_{R}} \rangle \\
\vert\downarrow_{R}\rangle &\eq& \sqrt{y}\ \vert {0_{R}} \rangle - \sqrt{x}\ \vert {1_{R}} \rangle
\end{eqnarray*}

and in this new basis the operator is

\begin{equation}
n_R \rightarrow {\mathbf M}_R = x{\cal P}^\uparrow_R + y{\cal P}^\downarrow_R + \sqrt{xy} \ ({\cal T}^{\uparrow\downarrow}_R 
+ {\cal T}^{\downarrow\uparrow}_R)
\label{eqd}
\end{equation}

These two vectors span the space $\phi_R$. The full configuration space $\Phi$ = $\prod^\otimes_R\ \phi_R$ is then 
spanned by vectors of the form $\vert\uparrow\uparrow\downarrow\uparrow\downarrow\ldots\rangle$.
These configurations may be labeled by the sequence of sites $\{{\cal C}\}$ 
 at which we have a $\downarrow$. For example, for the state just quoted  $\{{\cal C}\}$
= $\vert\{3,5,\ldots\}\rangle$. This sequence is called the {\sl cardinality sequence}.
If we define the configuration $\vert\uparrow\uparrow\ldots\uparrow\ldots\rangle$ as the {\sl average} or
 {\sl reference} configuration, then 
the {\sl cardinality sequence} of the {\sl reference} configuration is the null sequence 
$\{\emptyset\}$.

The augmented space theorem \cite{Am} states that

\be 
\ll A(\{ n_{R}\}) \gg \eq < \{\emptyset\}\vert \tilde{{\mbf A}}\vert \{\emptyset\}> 
\label{eqe}
\ee

\n where 

\[ \tilde{\mbf A}(\{{\bf M_R}\}) \eq \int \ldots \int A(\{\lambda_{R}\})\ \prod d{\mbf P}(\lambda_{R}) \]

\n {\bf P}($\lambda_{R}$) is the spectral density of the self-adjoint operator ${{\mathbf M}}_{R}$. 

Applying this idea we may obtain exact expressions for configuration averages of Green matrices (in angular momentum space) both in the real and reciprocal space representations
\begin{eqnarray}
\ll \o{G}(R,R,z)\gg \ = \ \langle R\otimes\{\emptyset\} |{(z\tilde\o{I}- \tilde\o{H}^{(2)})}^{-1}
|R\otimes\{\emptyset\} \rangle
\label{eqf}
\end{eqnarray}
\begin{eqnarray}
\ll \o{G}(\k,z)\gg \ = \ \langle \k\otimes\{\emptyset\} |{(z\tilde\o{I}- \tilde\o{H}^{(2)})}^{-1}
|\k\otimes\{\emptyset\} \rangle
\label{eqg}
\end{eqnarray}

where {\bf G} and {\bf H}$^{(2)}$ are operators which are matrices in angular momentum space, and 
the augmented k-space basis $|\k,L\otimes\{\emptyset\} \rangle$ has the form

\[ (1/\sqrt{N})\sum_R \mbox{exp}(-i\k\cdot R)|R,L\otimes \{\emptyset\}\rangle \]

The augmented space Hamiltonian $\tilde\o{H}^{(2)}$ is constructed from the TB-LMTO
Hamiltonian $\o{H}^{(2)}$ by replacing each random variable $n_R$ by  operators ${\mathbf M}_R$.
It is an operator in the augmented space 
$\Psi$ = ${\cal H} \otimes \Phi$. The ASF maps 
a disordered Hamiltonian described in a Hilbert space ${\cal H}$ onto an ordered Hamiltonian 
in an enlarged space $\Psi$, where the space $\Psi$ is constructed as the outer product of the
space ${\cal H}$ and 
configuration space $\Phi$ of the random variables of 
the disordered Hamiltonian. 
 The  configuration space $\Phi$ is of rank 2$^{N}$ if there are $N$ muffin-tin 
spheres in the system. Another way of looking at $\tilde\o{H}^{(2)}$ is to note
that it is the {\sl collection} of all possible Hamiltonians for all possible
configurations of the system.

A little mathematics yields the following :
\begin{eqnarray}
\ll {\mathbf G(z)}\gg &=& \langle {\mathbf 1}\vert \left(\tilde {\mathbf A}+\tilde {\mathbf B}+\tilde \mathbf{F}
-\tilde\mathbf{ S}+(\tilde\mathbf{ J}+\tilde \mathbf{S}) \ \tilde\mathbf{ o} \ (\tilde \mathbf{J}+\tilde\mathbf{ S}) \right)^{-1} 
\vert {\mathbf 1} \rangle. 
\label{eqh}
\end{eqnarray}

\n for real space calculations :

\[\vert \mathbf {1} \rangle  = \frac{\mathbf{A}(\mathbf{\Delta}^{-1/2})}{[\mathbf{A}(\mathbf{\Delta}^{-1})]^{1/2}}\ |R\otimes\{\emptyset\} \rangle 
+\frac{\mathbf{F}(\mathbf{\Delta}^{-1/2})}{[\mathbf{A}(\mathbf{\Delta}^{-1})]^{1/2}}\ |R\otimes\{R\}\rangle.  \]

\n or for reciprocal space calculations : 

\[\vert \hat{\mathbf 1}\rangle = \frac{\mathbf{A}(\mathbf{\Delta}^{-1/2})} {[\mathbf{A}(\mathbf{\Delta}^{-1})]^{1/2}}\ |\k\otimes\{\emptyset\} 
\rangle+\frac{\mathbf{F}(\mathbf{\Delta}^{-1/2})}{[\mathbf{A}(\mathbf{\Delta}^{-1})]^{1/2}}\ |\k\otimes\{R\}\rangle.  \]

where 

\[ \tilde \o{K}  =  \sum_{R} \left\{ \o{K}\left(\rule{0mm}{3mm} (z\mathbf{I}-\mathbf{C})\mathbf{\Delta}^{-1}\right)/\mathbf{A}(\mathbf{\Delta}^{-1}) \right \}\enskip {\cal O}_K  \] 

and $\o{K}$ may be {\bf A}, {\bf B} or {\bf F}, while the operators :

\[ {\cal O}_A = {\cal P}_R\otimes {\cal I}\quad {\cal O}_B = {\cal P}_{R}\otimes {\cal P}^{\downarrow}_{R}\quad
{\cal O}_F = {\cal P}_{R}\otimes \left\{ {\cal T}_{R}^{\uparrow\downarrow} + {\cal
T}_{R}^{\downarrow\uparrow} \right\}\]

Moreover,  $\tilde {\bf J}=\tilde{\bf  J}_A+\tilde {\bf J}_B+\tilde {\bf J}_F$ and $\tilde{\bf o}=\tilde {\bf o}_A+\tilde {\bf o}_B+\tilde {\bf o}_F$ where :

\begin{eqnarray}
\tilde \o{J}_K & = & \sum_{R} \left\{\o{K}\left(\rule{0mm}{3mm} (\o{C}-\o{E}_{\nu })\o{\Delta}^{-1}\right)/\o{A}(\o{\Delta}^{-1})\right\}\enskip 
{\cal O}_K \nonumber\\
\tilde \o{o}_K & = & \sum_{R} \left\{\rule{0mm}{3mm}\o{K}(\o{o}\o{\Delta}) \ \o{A}(\o{\Delta}^{-1})\right\} \enskip 
{\cal O}_K 
\label{eqi}
\end{eqnarray}
 
For any diagonal (in angular momentum space) operator {\bf V} : 
\begin{eqnarray*}
{\mathbf A}({\mathbf V}) =  ( x\ V^A_L + y\ V^B_L)\ \delta_{LL'}, \\
{\mathbf B}({\mathbf V}) =  (y-x)\ (V^A_L -  V^B_L)\ \delta_{LL'}, \\
{\mathbf F}({\mathbf V}) =  \sqrt{xy}\ (V^A_L -  V^B_L)\ \delta_{LL'}. 
\end{eqnarray*}

In case there is no off-diagonal disorder due to local lattice distortion because of size mismatch :

\[ \tilde\o{S} = \sum_{R}\sum_{R'} \o{A}(\o{\Delta}^{-1})^{-1/2}\ \o{S}_{RR'} \ \o{A}(\o{\Delta}^{-1})^{-1/2} \enskip
{\cal T}_{RR'}\otimes {\cal I}. \]

The equation (\ref{eqh}) is now exactly in the form in which recursion method may be applied. For ordinary 
recursion $\vert {\mathbf 1}\rangle$ is labeled by a particular $L$ and has a representation as
a $1\times N_{\mathrm{max}}$ column vector, with $N_{\mathrm{max}}$ being the size in augmented space.
But in the case of block recursion, we deal with $L\times L'$ matrix and the length of
$\vert {\mathbf 1}\rangle$ is $L\times L'\times N_{\mathrm{max}}$. We should note that at  this point
the above expression for the averaged $\ll G_{RL,RL'}(z)\gg$ or $\ll G_{LL'}({\bf k},z)\gg$  is {\sl exact}.

It is important to note that the operators $\tilde {\mathbf A},\ \tilde{\mathbf B},\ \tilde{\mathbf F},\ 
\tilde{\bf J}_A,\ \tilde{\bf J}_B,\ \tilde{\bf J}_F,\ \tilde{\bf o}_A,\ 
\tilde{\bf o}_B$ and $\tilde{\bf o}_F$ are all projection operators
in real space ({\sl i.e} unit operators in {\bf k-} space) and acts on an augmented space basis only to 
change the configuration part ({\sl i.e.} the cardinality sequence $\{{\cal C}\}$).
\begin{eqnarray*}
\fl {\tilde {\mathbf A}} \vert\vert \{ {\cal C} \} \rangle = A_{1} \vert\vert \{ {\cal C} \}\rangle \quad
& {\tilde {\mathbf B}} \vert\vert \{ {\cal C} \} \rangle = A_{2} \vert\vert \{ {\cal C} \} \rangle \ \delta( R \in \{ {\cal C} \})\quad
& {\tilde {\mathbf F}} \vert\vert \{ {\cal C} \} \rangle = A_{3} \vert\vert \{ {\cal C} \pm R \} \rangle\\
\fl {\tilde{\mathbf J}_A} \vert\vert \{ {\cal C} \} \rangle = J_{1} \vert\vert \{ {\cal C} \}\rangle \quad
& {\tilde {\mathbf J}_B} \vert\vert \{ {\cal C} \} \rangle = J_{2} \vert\vert \{ {\cal C} \} \rangle \ \delta( R \in \{ {\cal C} \})\quad
& {\tilde {\mathbf J}_F} \vert\vert \{ {\cal C} \} \rangle = J_{3} \vert\vert \{ {\cal C} \pm R \} \rangle\\
\fl {\tilde{\mathbf o}_A} \vert\vert \{ {\cal C} \} \rangle = o_{1} \vert\vert \{ {\cal C} \}\rangle \quad
& {\tilde {\mathbf o}_B} \vert\vert \{ {\cal C} \} \rangle = o_{2} \vert\vert \{ {\cal C} \} \rangle \ \delta( R \in \{ {\cal C} \})\quad
& {\tilde {\mathbf o}_F} \vert\vert \{ {\cal C} \} \rangle = o_{3} \vert\vert \{ {\cal C} \pm R \} \rangle
\end{eqnarray*}

The coefficients $A_{1} - A_{3},\ J_1 - J_3$ and $o_1 - o_3$  have been expressed in equation (\ref{eqi}). 
The remaining operator $\tilde{\bf S}$ is off-diagonal in real space, but diagonal in {\bf k}-space. 

In the real space representation:

\[\tilde{\mathbf{S}}\ \vert R,\{{\cal C}\}\rangle\ =\ \sum_\chi\  {\mathbf S}(\chi)\ \vert R+\chi,\{
{\cal C}\}\rangle. \]
Here $\chi$'s are the nearest neighbour vectors. The operator ${\tilde\mathbf S}$ shifts the real-space site to its
nearest neighbour position by the vector $\chi$ and configuration space remain unchanged. However, in the reciprocal 
space representation, it  acts on an augmented space only to change the configuration part:

\[\tilde{\mathbf S}\vert\vert \{ {\cal C} \} \rangle = \sum_{\chi} \exp{(-\imath {\bf k}.\chi) }\ {\mathbf S}(\chi)\ 
\vert\vert \{ {\cal C} - \chi \} \rangle. \]

Here the operator rigidly shifts the entire down-spin configuration by the vector $\chi$. The operation of the
effective Hamiltonian is thus entirely in the configuration space.
 
As long as we wish to obtain only the diagonal elements of the Green function, as is required for
the local density of states or the spectral densities, the ordinary recursion as described by Haydock \etal \cite{hhk} suffices.  However, for response functions we shall need the full Green matrix. This will be described in
the following section.

\subsection{The configuration averaged current-current correlation function}

We shall quote here the result for the dominant term in the configuration averaged current-current correlation function reported by us earlier \cite{opt}. This correlation function is directly related to the optical conductivity. We note that the expression involves the full Green matrix in angular momentum space and not only its diagonal elements.

\n The expression for correlation function is
\begin{eqnarray}  
\fl \ll {\cal S}(z_1,z_2)\gg \eq \!\!\!\sumk \mbox{Tr}\left[\rule{0mm}{4mm}
\mbf{J}^{\mathrm{eff}}(\k, z_1,z_2)\ll \mbf{G}^v(\k,z_1)\gg
\mbf{J}^{\mathrm{eff}}(\k,z_1,z_2)^{\dagger}\ll \mbf{G}^c(\k,z_2)\gg\right] \nonumber \\
\label{eqj}
\end{eqnarray}

\n and the renormalized current term is given by :
\begin{eqnarray}
\fl \mbf{J}^{\mathrm{eff}}(\k,z_1,z_2) &=&  \ll {\bf j}(\k)\gg + 2\left[\rule{0mm}{5mm} \mbf{\Sigma}(\k,z_2)\ \mbf{f}(z_2)\
{\bf j}^{(1)}(\k) \pls {\bf j}^{(1)}(\k)\ \mbf{f}(z_1)\ \Sigma(\k,z_1)\right] \nonumber \\
&& \qquad\qquad \pls \Sigma(\k,z_2)\ \mbf{f}(z_2)\ {\bf j}^{(2)}(\k)\ \mbf{f}(z_1)\ \Sigma(\k,z_1)
\label{eqk}
\end{eqnarray}
\n where
\begin{eqnarray*}
\mbf{f}(z)\eq  f_{LL'}(z) & \eq  \rule{0mm}{6mm}\left\langle \frac{1}{\Delta_{L}}\right\rangle \left[ \frac{C^A_L}{\Delta^A_L} - \frac{C^B_L}{\Delta^B_L}
-z \left( \frac{1}{\Delta^A_L}-\frac{1}{\Delta^B_L}\right)\right]^{-1}\ \delta_{LL'} \\
\end{eqnarray*}
\n and
\[ {\mathbf \Sigma}\ =\ {\mathbf g}^{-1}\ -\ {\mathbf G}^{-1}\]

\n{\bf g} is the virtual crystal Green function.
The interested reader is referred to an earlier work \cite{opt} which derives these expressions in some detail. 
The main point in setting these equations out,  is to note
that in such calculations one needs the {\sl full} Green matrix in angular momentum space. This
is the main motivation for this work.

\subsection{Setting up the Block recursion}
\begin{figure}[h]
\centering
\epsfxsize=2in\epsfysize=1.9in \epsfbox{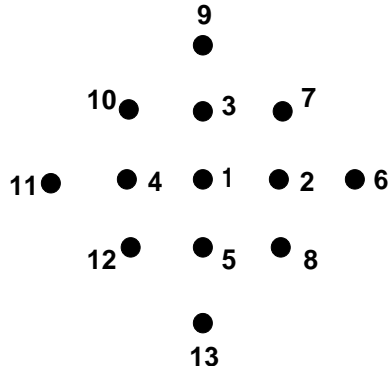}
\caption{Systematic discrete numbering of nearest neighbour map on a square lattice.}
\label{fig1}
\end{figure}

The first step in setting up the Block recursion procedure is to 
 systematically renumber the real-space basis with integers. An example on a square lattice is shown in \fref{fig1}.

The nearest neighbour map is now generated by a systematic numbering of the states in augmented space as follows : 

{\sl (A) Real space formulation :}
We start with numbering $\vert R,\{\emptyset\}\rangle $ as 1, and then recursively generate the neighbours by acting on the states by $\tilde{\bf S}$ and $\tilde{\mathbf F}$. Let us take an example of a square lattice :

\begin{enumerate}
\item  $\tilde{\bf S}$ acting on $\vert 1,\{\emptyset\}\rangle \equiv \vert 1\rangle $ gives
four new neighbours $\vert 2,\{\emptyset\}\rangle\ \ldots \ \vert 5,\{\emptyset\}\rangle$.
 The four real space neighbours of $\vert 1\rangle $ are then $\vert 2\rangle $,\ $\vert 3\rangle $,\ $\vert 4\rangle $ and $\vert 5\rangle $.
\item $\tilde{\bf F}$ acting on $\vert 1,\{\emptyset\}\rangle \equiv \vert 1\rangle $ is
$\vert 1,\{ 1\}\rangle$. This we number $\vert 6\rangle$.
\item $\tilde{\bf S}$ acting on $\vert 2,\{\emptyset\}\rangle \equiv \vert 2\rangle $ gives :
$\vert 6,\{\emptyset\}\rangle$, $\vert 7,\{\emptyset\}\rangle$, $\vert 1, \{\emptyset\}\rangle$ and $\vert 8,\{\emptyset\}\rangle$. These
we number $\vert 7\rangle$,  $\vert 8\rangle$, $\vert 1\rangle$  and $\vert 9\rangle$.
\item $\tilde{\bf F}$ acting on $\vert 2,\{\emptyset\}\rangle \equiv \vert 2\rangle $ is
$\vert 2,\{2\}\rangle \equiv \vert 10\rangle$.
\end{enumerate}

We proceed exactly as above and finally obtain the nearest neighbour map matrix : the
$n$-th column of whose $m$-th row gives the $n$-th neighbour of $m$. We show below
the initial part of the nearest neighbour map for the above example :

\begin{center}
$\left(
\begin{array}{ccccc}
2 & 3 & 4 & 5 & 6 \\
7 & 8 & 1 & 9 & 10 \\
8 & 11 & 12 & 1 & 13 \\
\multicolumn{5}{c}{\cdots\cdots\cdots\cdots\cdots}\\
\end{array}\right)$
\end{center}

The equivalences are :

\begin{tabular}{lccccccccc}
Augmented-space element & $\Rightarrow$ & 1\{$\emptyset$\} &  2\{$\emptyset$\} & 3\{$\emptyset$\} & 4\{$\emptyset$\} & 5\{$\emptyset$\} &  1\{1\} & 6\{$\emptyset$\} &  \\
Discrete numbering & $\Rightarrow$ & 1 & 2 & 3 & 4 & 5 & 6 & 7  \\
 & & & & &  \\
Augmented-space element & $\Rightarrow$ & 7\{$\emptyset$\} & 8\{$\emptyset$\} &  2\{2\} & 9\{$\emptyset$\} & 10\{$\emptyset$\} & 3\{3\} &    &      \\
Discrete numbering & $\Rightarrow$ & 8 & 9 & 10 & 11 & 12 & 13 &  &    \\
\end{tabular}
\vskip 0.3cm

{\sl (B) Reciprocal space formulation :}\\
In reciprocal space the procedure is even simpler, since the operators act {\sl only}
on the configuration part of the space. As before, we start with numbering $\vert k,\{\emptyset\}\rangle$ as $1$ and then recursively generate the neighbours by acting on the states successively by $\tilde{\bf S}$
and $\tilde{\bf F}$. Let us take the example of the square lattice :

\begin{enumerate}
\item  $\tilde{\bf S}$ acting on $\vert\{\emptyset\}\rangle \equiv \vert 1\rangle $ leaves it unchanged. The four neighbours of $\vert 1\rangle $ are then $\vert 1\rangle $,\ $\vert 1\rangle $,\ $\vert 1\rangle $ and $\vert 1\rangle $.
\item $\tilde{\bf F}$ acting on $\vert\{\emptyset\}\rangle \equiv \vert 1\rangle $ is
$\vert \{ 1\}\rangle$. This we number $\vert 2\rangle$.
\item $\tilde{\bf S}$ acting on $\vert\{1\}\rangle \equiv \vert 2\rangle $ gives :
$\vert\{2\}\rangle$, $\vert\{3\}\rangle$, $\vert\{4\}\rangle$ and $\vert\{5\}\rangle$. These
we number $\vert 3\rangle$,  $\vert 4\rangle$, $\vert 5\rangle$  and $\vert 6\rangle$.
\item $\tilde{\bf F}$ acting on $\vert\{1\}\rangle \equiv \vert 2\rangle $ is
$\vert\{\emptyset\}\rangle \equiv \vert 1\rangle$.
\end{enumerate}

We proceed  as before and  obtain the nearest neighbour map matrix. 
 We again show below
the initial part of the nearest neighbour map in reciprocal space :

\begin{center}
$\left(
\begin{array}{ccccc}
1 & 1 & 1 & 1 & 2 \\
3 & 4 & 5 & 6 & 1 \\
7 & 8 & 2 & 9 & 10 \\
\multicolumn{5}{c}{\cdots\cdots\cdots\cdots\cdots}\\
\end{array}\right)$
\end{center}

Now the equivalences are :
\begin{center}
\begin{tabular}{lccccccccccc}
Cardinality sequence & $\Rightarrow$ & \{$\emptyset$\} & \{1\} & \{2\} & \{3\} & \{4\} 
 & \{5\} & \{6\} & \{7\} &  \{8\} & \{1, 2\}  \\
Discrete numbering &  $\Rightarrow$ & 1 & 2 & 3 & 4 & 5  & 
 6 & 7 & 8 &  9 & 10  \\
\end{tabular}
\end{center}
{\sl (C) The Block recursion :}\\

We now go over to a matrix basis of the form : $\{\Phi^{(n)}_{J,LL'}\}$, where $J$ is the 
the discrete labeling of the augmented space states and $L,L'$ labels the angular momenta.
The inner product of such basis is defined by :

\[\left( \Phi^{(n)}, \Phi^{(m)}\right) \ =\ \sum_J\sum_{L''}\ \Phi^{(n)\dagger}_{LL'',J}\ 
\Phi^{(m)}_{J,L''L'}\ =\ N^{nm}_{LL'} \]

For a real space calculations on a lattice with $Z$ nearest neighbours, we  start the recursion with :

\[ \Phi^{(1)}_{J,LL'}\ =\ {\mathbf W}^{(1)}_{LL'}\ \delta_{J,1}\ +\ {\mathbf W}^{(2)}_{LL'}\ \delta_{J,Z+1}\]
While for a reciprocal space calculation we start with :

\[ \Phi^{(1)}_{J,LL'}\ =\ {\mathbf W}^{(1)}_{LL'}\ \delta_{J,1}\ +\ {\mathbf W}^{(2)}_{LL'}\ \delta_{J,2}\]

\n where

\begin{equation}
{\mathbf W}^{(1)}_{LL'} \ =\ \frac{A(\Delta_L^{-1/2})}{\left[A(\Delta_L^{-1})\right]^{1/2}}\ \delta_{LL'} \quad 
{\mathbf W}^{(2)}_{LL'} \ =\ \frac{F(\Delta_L^{-1/2})}{\left[A(\Delta_L^{-1})\right]^{1/2}}\ \delta_{LL'} 
\label{eql}
\end{equation}

The remaining terms in the basis are recursively obtained from :

\begin{eqnarray*}
\fl \sum_{L''}\Phi^{(2)}_{J,LL''}\ B^{(2)\dagger}_{L''L'}&\ =\ &\sum_{J'}\sum_{L''} H_{JL,J'L''}\ 
\Phi^{(1)}_{J',L''L'}\ -\ \sum_{L''}\Phi^{(1)}_{J,LL''}\ A^{(1)}_{L''L'} \\
\fl\sum_{L''} \Phi^{(n+1)}_{J,LL''}\ B^{(n+1)\dagger}_{L''L'}&\ =\ &\sum_{J'}\sum_{L''} H_{JL,J'L''}\ 
\Phi^{(n)}_{J',L''L'}\ -\ \sum_{L''}\Phi^{(n)}_{J,LL''}\ A^{(n)}_{L''L'}\ -\ \sum_{L''}\Phi^{(n-1)}_{J,LL''}\
B^{(n)}_{L''L'}
\end{eqnarray*}

Orthogonalization of the basis gives :

\begin{equation} 
\fl\sum_{J}\sum_{L''}\sum_{J'}\sum_{L'''}\ \Phi^{(n)\dagger}_{LL'',J}\ H_{JL'',J'L'''}\ \Phi^{(n)}_{J',L'''L'}
\ =\ \sum_{L''}\ N^{nn}_{LL''}\ A^{(n)}_{L''L'} 
\label{eqm}
\end{equation}

In matrix notation, where matrices are in angular momentum space :

\begin{equation} 
{\mathbf A}^{(n)} \ =\ \left({\mathbf N}^{nn}\rule{0mm}{4mm}\right)^{-1}\ \sum_{J}\sum_{J'}\ {\mathbf \Phi}^{(n)\dagger}_J\ 
{\mathbf H}_{JJ'}\ {\mathbf \Phi}^{(n)}_{J'} 
\label{eqn} 
\end{equation}

Next, we note that we had started with $J_{\mathrm{max}}\times L_{\mathrm{max}}^2$ orthogonal basis set. The above procedure merely gives $J_{\mathrm{max}}$ basis sets. We still have orthogonality conditions among
the various columns of $\Phi^{(n)}_{J,LL'}$. In order to impose these conditions consider :

\[\fl \Psi_{J,LL'}\ =\ \sum_{J'}\sum_{L''} H_{JL,J'L''}\ 
\Phi^{(n)}_{J',L''L'}\ -\ \sum_{L''}\Phi^{(n)}_{J,LL''}\ A^{(n)}_{L''L'}\ -\ \sum_{L''}\Phi^{(n-1)}_{J,LL''}\
B^{(n)}_{L''L'}\]

Construct $L_{\mathrm{max}}$ column vectors  $\phi_{JL}^{(L')}$ out of the $L_{\mathrm{max}}$ columns of $\Psi_{J,LL'}$ and set about to Gram-Schmidt orthonormalizing the set :

\begin{eqnarray}
\fl\phi^{(1)}_{LJ} \ =\ B_{11}\psi^{(1)}_{LJ}\quad\Rightarrow\quad
B_{11}^2 \ =\ \sum_{LJ}\phi^{(1)}_{JL} \phi^{(1)}_{LJ}\nonumber\\
\fl\phi^{(2)}_{LJ} \ =\ B_{21}\psi^{(1)}_{LJ} + B_{22}\psi^{(2)}_{LJ}\ \ \Rightarrow \ \
B_{21} \ =\ \sum_{LJ} \psi^{(1)}_{JL}\phi^{(2)}_{LJ} \ ; \ 
B_{22}^2 \ =\  \sum_{LJ}\phi^{(1)}_{JL} \phi^{(1)}_{LJ} - B_{21}^2\nonumber\\
\ldots\quad\ldots\quad\ldots\quad\ldots\quad\ldots\quad\ldots\quad\ldots\quad\ldots\quad\ldots\nonumber\\
\fl\phi^{(n)}_{JL} \ =\ \sum_{k=1}^{n}\ B_{nk} \phi^{(k)}_{LJ}\ \ \Rightarrow \ \
B_{nk}\ =\ \sum_{LJ} \psi^{(k)}_{JL}\phi^{(n)}_{LJ} \quad (k < n) \ ; \
B_{nn}^2 \ =\ \sum_{LJ} \phi^{(n)}_{JL} \phi^{(n)}_{LJ} - \sum_{k=1}^{n-1} B^2_{nk}\nonumber\\
\label{eqo}
\end{eqnarray}

We may now  construct  $\Phi^{(n+1)}_{J,LL'}$ out of $\psi_{JL}^{L'}$ and 
note that   $B_{nk}$ is indeed the matrix ${\bf B}^{(n+1)}$ we are looking for.

The equations (\ref{eqn}-\ref{eqo}) show that we may calculate the matrices $\{{\bf A}^{(n)}, {\bf B}^{(n+1)}\}$ recursively, noting that ${\bf B}^{(1)}\ =\ {\bf I}$ and ${\bf B}^{(0)}\ =\ {\bf 0}$.
In this new basis, the Hamiltonian is  {\sl block tri-diagonal} and the Green matrix can be written as follows :

\begin{eqnarray}
&&{\mathbf G}_{(n)} \ = \ \left[\rule{0mm}{4mm} E\ {\mathbf I}- {\mathbf A}^{(n)}- {\mathbf B}^{(n+1)\dagger}\ {\mathbf G}_{(n+1)}\ 
{\mathbf B}^{(n+1)}\right]^{-1} \nonumber\\
&&\phantom{x} \nonumber\\
&&\ll {\mathbf G}\gg \ = \ {\mathbf G}_{(1)}
\label{eqp}
\end{eqnarray}

\section{Model Calculations}

{\sl (A) Model on a square lattice :} We shall first apply our methodology to a two-band model on a square lattice with 
the Hamiltonian~:

\begin{eqnarray}
 H^A\ =\ \sum_R\ \left( \begin{array}{cc}
                1.0 & 0.0 \\
                0.0 & 1.0\end{array}\right) \ P_R + 
         \sum_R\sum_{R'}\ \left(\begin{array}{rr}
                        -2.0 & -0.2 \\
	                -0.2 & -0.5 \end{array}\right)\ T_{RR'}\nonumber \\
H^B \ =\ \sum_R\             \left(\begin{array}{cc}
                0.1 & 0.0\\
                0.0 & 0.1 \end{array}\right)\ P_R +
         \sum_R\sum_{R'}\ \left(\begin{array}{rr}
                        -2.0 & -0.2 \\
	                -0.2 & -0.5 \end{array}\right)\ T_{RR'}
\label{eqq} 
\end{eqnarray}

The concentration is taken to be $x=0.5$.

\begin{figure}
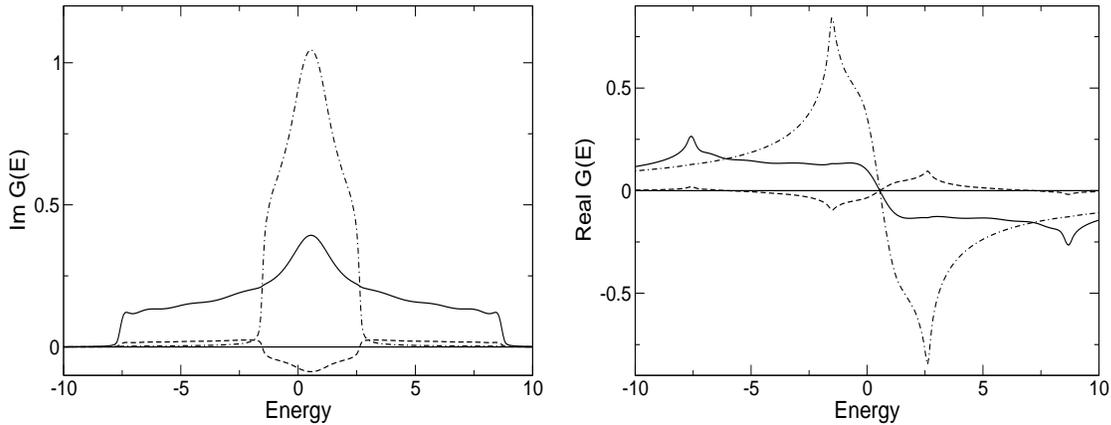

\centering
\vskip 0.8cm
\epsfxsize=2.8in\epsfysize=2.2in\epsfbox{fig2a.eps}\quad
\epsfxsize=2.8in\epsfysize=2.2in\epsfbox{fig2b.eps}
\caption{The real and imaginary parts of the Green matrix for a $2\times 2$ Hamiltonian
model. The full lines refer to {\bf G}$_{11}$, the dashed-dotted line to {\bf G}$_{22}$ and
dashed lines to {\bf G}$_{12}$.}
\label{fig2}
\end{figure}

We have carried out block recursion for $N$=10  levels. The termination was carried out as suggested earlier by Godin 
and  Haydock \cite{gh2} : we  put $\{{\mbf A}^{(n)}, {\mbf B}^{(n+1)}\}$ = $\{{\mbf A}^{(N)}, {\mbf B}^{(N+1)}\}$ for 
$n=N+1,\ldots, N_{\mathrm{max}}$ and take ${\mbf G}_{(N_{\mathrm{max}}+1)} = \left(\rule{0mm}{3mm}1/(E-i\delta)\right) {\mbf I}$. 
In order to get a smooth density of states we had taken $\delta = 0.01$ and $N_{\mathrm{max}} = 10000$. The elements of the
Green matrix is shown in \fref{fig2}. The imaginary part of the diagonal elements give the projected density of states. 
Herglotz properties of the diagonal parts give rise to positive definite density of states. The off-diagonal part is 
relatively small and is not Herglotz. The projected density of states
are symmetric, as is the imaginary part of the off-diagonal element. The real parts of the matrix elements are also shown. These are related to the imaginary parts by the Kramers-Kr\"onig relation.

\vskip 0.2cm
{\sl (B) The s-d model of a transition-noble metal alloy :} Levin and Ehrenreich \cite{le} and Gelatt and Ehrenreich \cite{ge} have introduced a simple two band model for transition-noble metal alloys. Physical effects like charge transfer between constituents  will usually differ for the $s$-$p$ conduction bands on one hand and the relatively narrow $d$-bands on the other. Their model
includes the conduction bands described together as a single band and the set of $d$-bands also described by a single band and their hybridization. The model also takes into account the large widths of the conduction bands and the relatively narrower widths of the $d$-bands. The following Hamiltonian has many (but not all) of the essential features :
\begin{eqnarray}
H \ =\ \sum_{R}\ \left( \begin{array}{rr}
                        \epsilon_s & \gamma \\
                        \gamma & \epsilon_d \end{array}\right)\ P_R +\sum_R\sum_{R'}
\ \left(\begin{array}{rr}
	t_s & 0 \\
         0 & t_d \end{array}\right)\ T_{RR'}
\label{eqr}
\end{eqnarray}

The dominant disorder is taken to be in the terms $\epsilon_s$ and $\epsilon_d$. The
hybridization is taken between states at the same site only and the hopping terms are
related by $t_s\ =\ \alpha t_d$. The sites $R$ vary over the sites of a face-centered cubic lattice. 
As a model case we have taken the parameters shown in the following table~:

\begin{table}[h]
\centering
\begin{tabular}{lrrrrr}\hline
Constituent & $\epsilon_s$ & $\epsilon_d$ & $\gamma$ & $\alpha$ & $t_d$ \\ \hline
$A$ & 1.5 & 1.5 & 0.2 & 4.0 & -0.5 \\
$B$ & 0.0 & 0.0 & 0.2 & 4.0 & -0.5 \\
\hline\end{tabular}
\caption{Parameters for our calculation for $A_{50} B_{50}$ alloy on a fcc lattice.}
\end{table}

{\sl (i) Calculations using real space block recursion :} \\

\begin{figure}
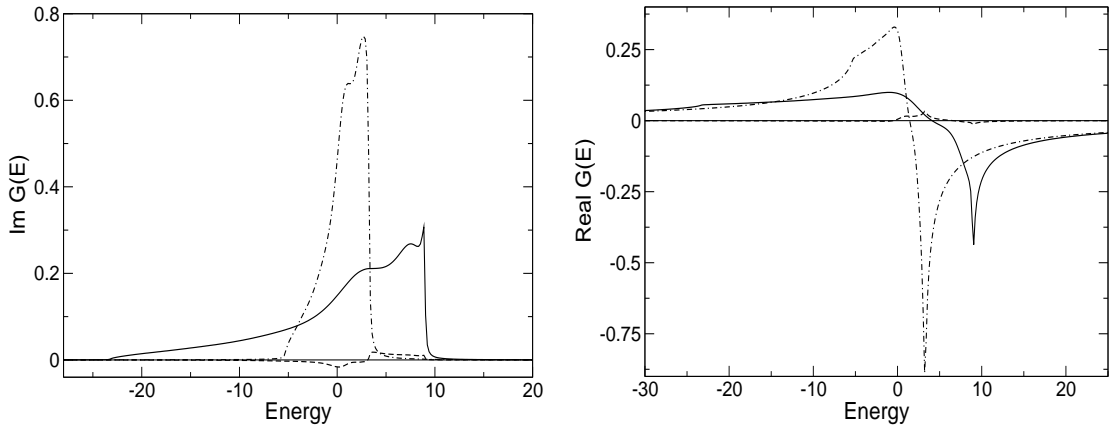

\centering
\vskip 0.6cm
\epsfxsize=2.8in\epsfysize=2.2in\epsfbox{fig3a.eps}\quad
\epsfxsize=2.8in\epsfysize=2.2in\epsfbox{fig3b.eps}
\caption{The real and imaginary parts of the Green matrix for the $s-d$ model on a fcc lattice. The calculations were 
done by a real space block recursion. The full lines refer to {\bf G}$_{11}$, the dashed-dotted line to {\bf G}$_{22}$ and
dashed lines to {\bf G}$_{12}$.}
\label{fig3}
\end{figure}

The \fref{fig3} shows the real and imaginary parts of the Green matrix. These calculation are carried out through a
real space block recursion technique. In general the qualitative features for the diagonal elements are similar to our 
square-lattice model. The main difference is that on a fcc lattice the partial density of states, related to the imaginary 
part of the diagonal elements of the Green matrix, are no longer symmetric about the band centre. Consequently, the 
behaviour of the off-diagonal element is quite different. The total density of states is given by : 
\[ \mbox{n}(E) = (1/\pi) \ \mbox{Im} \left(\rule{0mm}{4mm} G_{\rm {ss}}(E-i\delta)+ 5G_{\rm{dd}}(E-i\delta)\right). \]
If, for example,  the number of electrons in the constituents are 5 per atom per spin for $A$ and 5.5 per atom per spin for $B$, 
the position of the Fermi energy is given by :
\[\int_{-\infty}^{E_{\rm F}}\ dE\ {\mbox n}(E)\ =\ \langle {\rm n}_{\rm e}\rangle \ =\ x \ {\rm n}_A+(1-x) \ {\rm n}_B\ =\ 5.25 \]
where ${\rm n}_A$ and ${\rm n}_B$ are the number of valence electrons of $A$ and $B$ type of atom.

\begin{figure} 
\centering
\epsfxsize=6.1in \epsfysize=2.0in \rotatebox{0}{\epsfbox{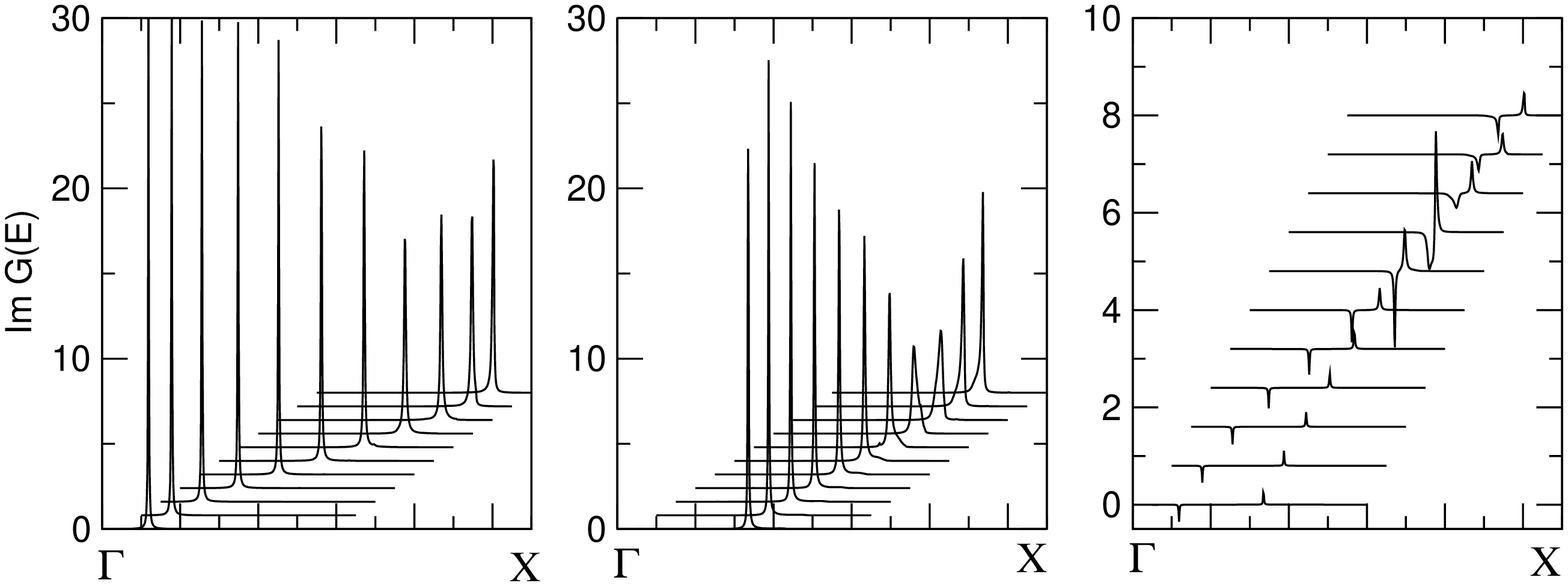}}
\caption{The matrix elements of {\bf G}({\bf k},E) on a fcc lattice : the 11 element (left),
the 22 element (middle), and the 12 element (right) along the $\Gamma$ to $X$ direction.} 
\label{fig4}
\end{figure}

{\sl (ii) Calculations using reciprocal space block recursion :} \\

We have carried out the block recursion in reciprocal space for the $s-d$ model. The Green matrix in reciprocal space 
${\bf G}(\k,E)$ is the factor that arises in our earlier expression for the configuration averaged
current-current correlation function. Its diagonal matrix element is related to the spectral functions for the 
different bands. In \fref{fig4} we show the spectral functions
along a given direction $\Gamma - X$ in reciprocal space. The imaginary part of the
off-diagonal matrix element is also shown in the figure. We note that the off-diagonal parts of the Green
matrix has antisymmetric structure in its peaks, while the diagonal matrix elements are positive (as they represent spectral functions).

\begin{figure}[b]
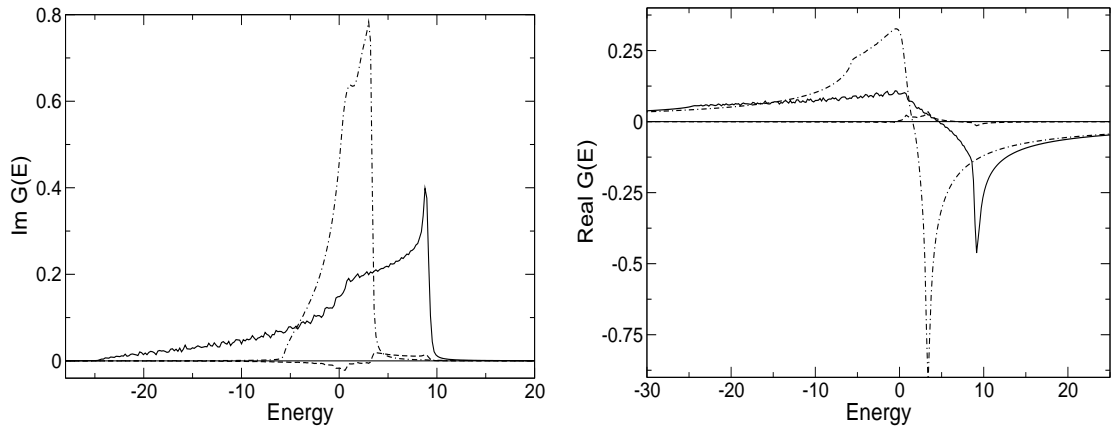

\centering
\vskip 0.2cm
\epsfxsize=2.8in\epsfysize=2.2in\epsfbox{fig5a.eps}\quad
\epsfxsize=2.8in\epsfysize=2.2in\epsfbox{fig5b.eps}
\caption{The real and imaginary parts of the Green matrix for the $s-d$ 
model on a fcc lattice. The calculations are done by a k-space block recursion followed by a 
reciprocal space integration \cite{rsi}.
 The full lines refer to {\bf G}$_{11}$, the dashed-dotted line to {\bf G}$_{22}$ and
dashed lines to {\bf G}$_{12}$.}
\label{fig5}
\end{figure}

Figure \ref{fig5} shows the Green matrix elements calculated starting from the ${\mathbf G}(\k,E)$ and carrying 
out a reciprocal space integration developed by us \cite{rsi} as a generalization of the tetrahedron method
proposed by Jepsen and Andersen \cite{andersen} for crystalline systems. The close comparison between the
figures \ref{fig3} and \ref{fig5} is a strong justification of the accuracy of the reciprocal space recursion.

\begin{figure}
\centering
\vskip 0.2cm
\epsfxsize=5.5in \epsfysize=4.2in \rotatebox{0}{\epsfbox{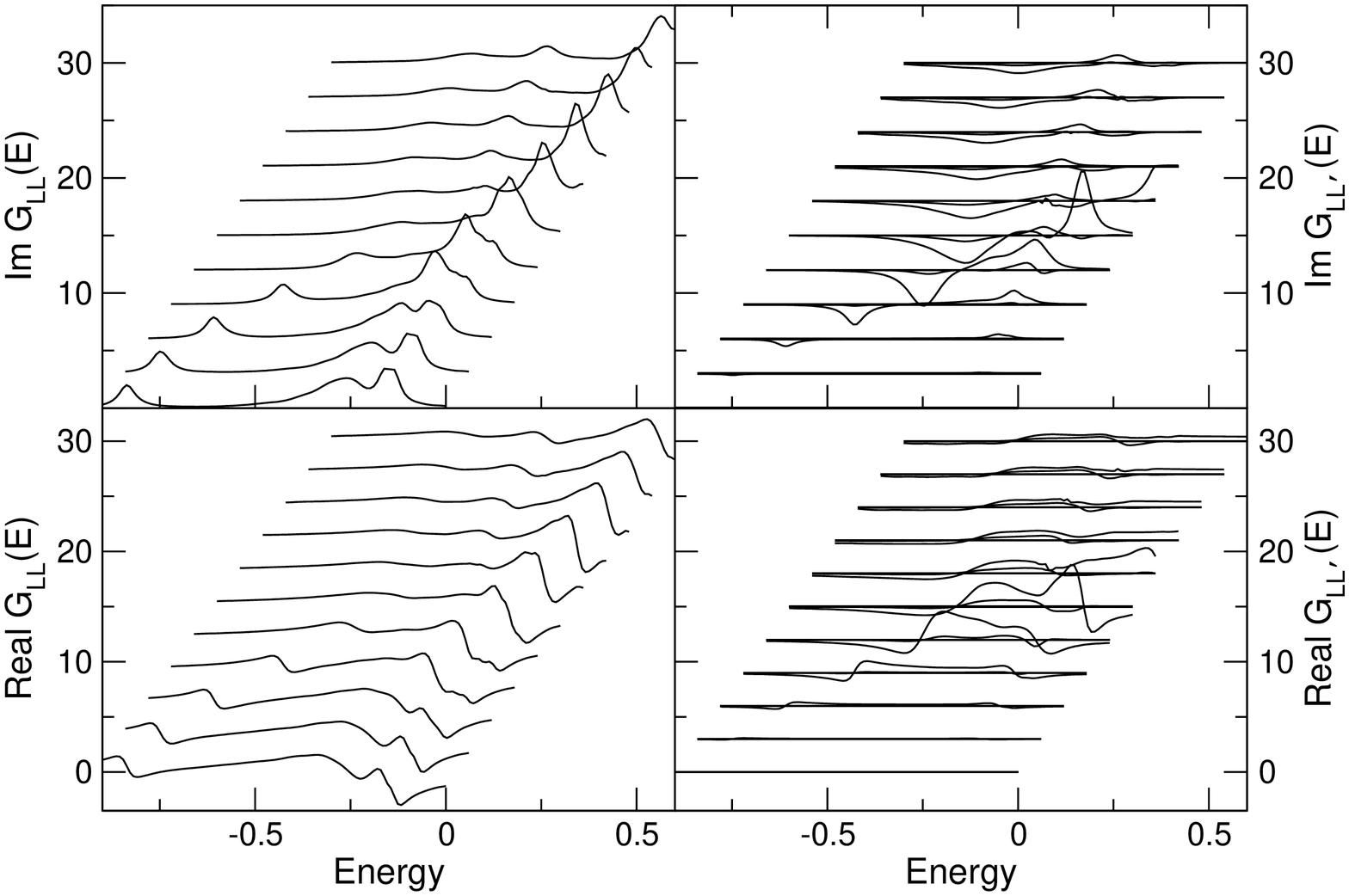}}
\caption{The imaginary and real parts of (1/9)Tr G$_{LL}(\k, E)$ (left column) and the imaginary 
and real parts of G$_{LL'}(E)$ (right column) for Ni$_{50}$Pt$_{50}$ alloy.}
\label{fig6}
\end{figure}

\section{Optical conductivity of Ni$_{50}$Pt$_{50}$ alloy} 

\begin{figure}
\centering
\vskip 0.2cm
\epsfxsize=4.5in \epsfysize=3.2in \rotatebox{0}{\epsfbox{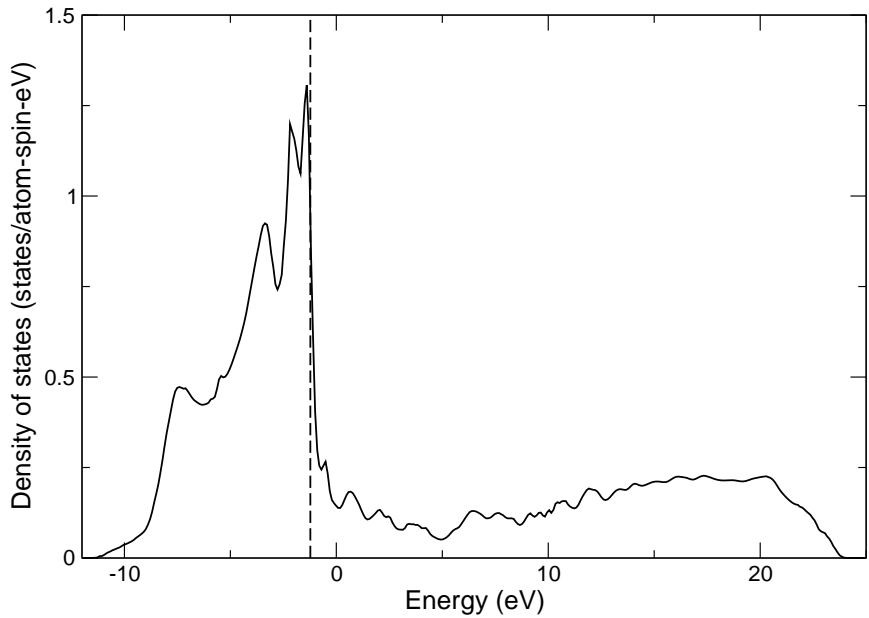}}
\caption{The density of states for a 50-50 NiPt alloy. The dashed line shows the Fermi energy.}
\label{fig7}
\end{figure}

We shall now report a calculation of the optical conductivity of a 50-50 NiPt alloy. The initial electronic structure 
calculations require input of the potential parameters from the pure Ni and Pt. This we have taken from a TB-LMTO 
calculation on the pure materials. We use these as an input for a self-consistent calculation of the electronic 
structure using our LDA-self consistent Augmented space recursion technique (ASR). We first calculate the Green matrix 
in reciprocal space. The results are shown in figure~\ref{fig6}. The imaginary part of the diagonal elements yield 
the spectral functions. The figure~\ref{fig6} also shows the off-diagonal matrix elements. Most of these elements are 
very small as compared with the spectral functions and are not positive definite as expected. The density of states 
for the alloy is shown in figure~\ref{fig7}. We obtain the density of states by integrating the spectral function over 
the Brillouin zone, using the tetrahedron method generalized for disordered alloys by us \cite{rsi}. The density of states
for NiPt is particularly difficult to reproduce accurately by real space recursion, because of the sharp structure
straddling the Fermi energy. However, the calculation in reciprocal space followed by Brilluoin zone integration 
reproduces the density of states in very good agreement with earlier work using KKR-CPA \cite{staunton}.

\vskip 0.2cm

The calculation of the optical conductivity involves the calculation of the four current terms $\o{j}^{AA}$,
 $\o{j}^{AB}$, $\o{j}^{BA}$ and $\o{j}^{BB}$. Ideally one should calculate these out of two atoms of the type AA, AB,
BA or BB embedded in the disordered alloy. However, as a first approximation we have obtained these from the pure 
metals and the ordered alloy. Having obtained the current terms we use the scattering methodology described in 
\cite{opt} to obtain the correlation function.

\begin{figure}[t]
\centering
\vskip 0.2cm
\epsfxsize=5.3in \epsfysize=3.0in \rotatebox{0}{\epsfbox{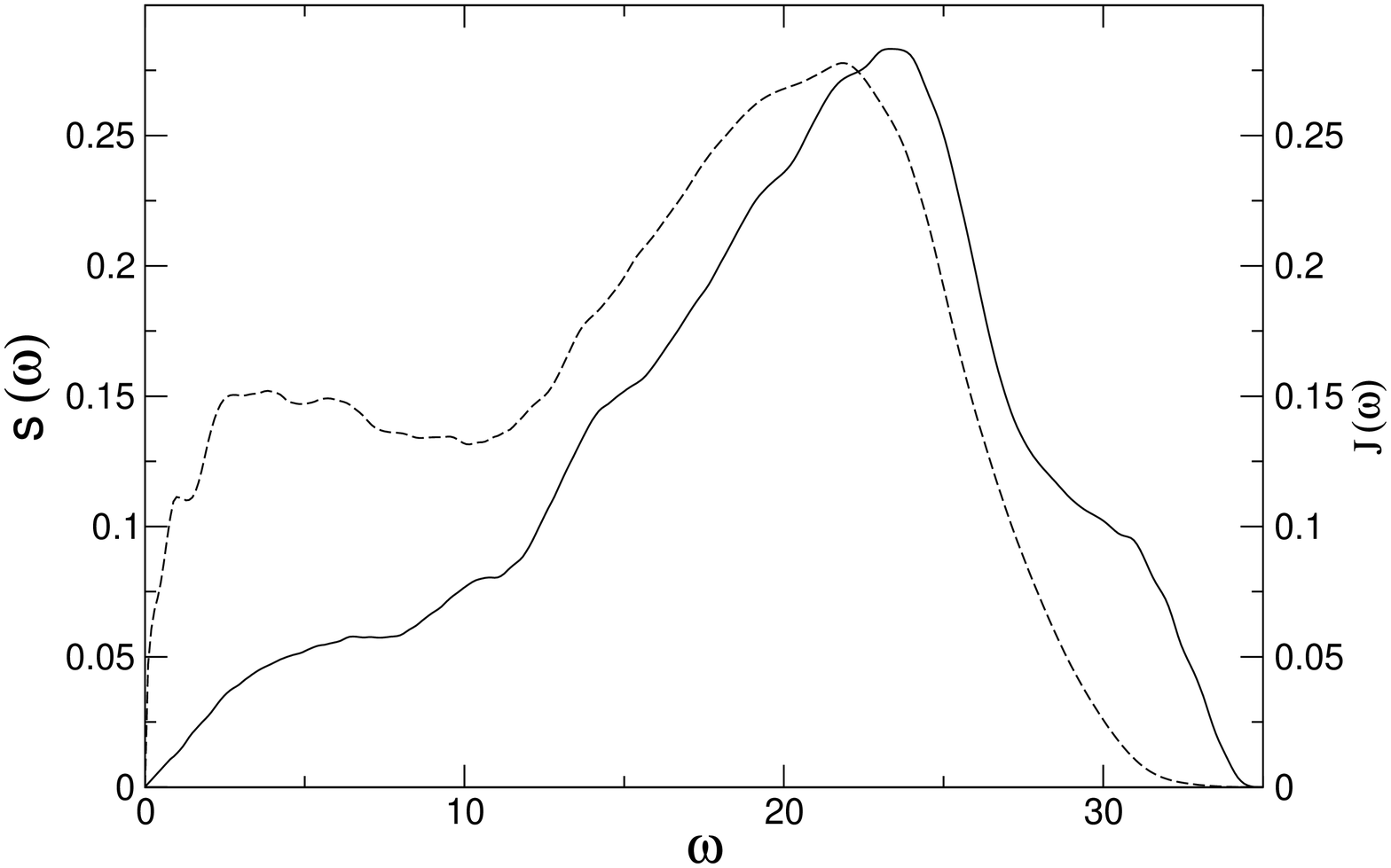}}
\caption{The joint density of states J($\omega$) (dashed line) and the
correlation function  $S(\omega)$ (solid line) for 50-50 NiPt alloys.}
\label{fig8}
\end{figure}

\vskip 0.2cm
In figure~\ref{fig8} we show the result for
the correlation function (full lines) for a 50-50 NiPt alloy :
\[ S(\omega) = \frac{1}{4} \int\ dE_1 \left[\rule{0mm}{4mm} S(E_1^+,E_2^+)+S(E_1^-,E_2^-)-S(E_1^+,E_2^-)-S(E_1^-,E_2^+)\right]. \] 
where $E_2 = E_1+\omega$, $E^{\pm} = E\pm i0^+$. For comparison we have also shown (as dashed lines) the joint density of states  for
the same alloy 
\[J(\omega) = \int\ dE\ {\mbox n}_{\mbox v}(E)\ {\mbox n}_{\mbox c}(E+\omega). \]
The former calculation requires the full Green matrices as well as the full matrix self-energies (matrices in the angular momentum 
space). We have carried out the calculations of these matrices via our block recursion technique. The joint density of states is 
modulated by the current terms. The main structures of the joint density of states are retained in the correlation function, but 
the relative heights of the principal peaks are modified. \\

\section{Conclusion}

In this communication we have described a block recursion in augmented space suitable for calculations of the Green matrices. 
The recursion is set up both with real space augmented by the configuration space of the alloy or the reciprocal space
augmented with the configuration space. For the latter case we have coupled it with a Brillouin zone integration scheme 
which is a generalization of the tetrahedron method developed earlier for crystalline systems. The Green matrices are
essential for the calculation of the response functions and effective current terms which are related to the self-energy 
matrices. We propose to use these techniques in our future applications.

\end{document}

%% file: mak.tex
\def\o#1{\mathbf{#1}}

\def\ket{\vert \vert  \{ \emptyset \} \rangle}
\def\ket2{\vert \vert \otimes \{ R \} \rangle}

\def\n{\noindent }
\def\mbf#1{{\mathbf {#1}}}

\def\eq{\ =\ }

\def\pls{\ +\ }
\def\be{\begin{equation}}
\def\ee{\end{equation}}
\def\<{[}
\def\>{]}

\def\k{{\mathbf k}}

\def\sumk{\int_{\rm{BZ}}\ \frac{d^3\k}{8\pi^3}\ }

\def\ket{\vert \vert  \{ \emptyset \} \rangle}
\def\ket2{\vert \vert \otimes \{ R \} \rangle}

\def\n{\noindent }
\def\mbf#1{{\mathbf {#1}}}

\def\eq{\ =\ }

\def\pls{\ +\ }
\def\be{\begin{equation}}
\def\ee{\end{equation}}
\def\<{[}
\def\>{]}

\def\k{{\mathbf k}}

\def\prb{{\sl Phys. Rev.} {\bf B }}

%% file: paper.bbl
\begin{thebibliography}{99}
\bibitem{her} A function of a complex variable f(z) is called Herglotz if : all singularities of f(z) 
lie on the real z axis and Im f(z)~=~$-$ sgn(Im z) for all z off the real axis.
\end{thebibliography}
